\PassOptionsToPackage{sort&compress}{natbib}

\documentclass[final,3p,times]{elsarticle}

\usepackage{booktabs}
\usepackage{ulem}
\usepackage{float}
\usepackage{siunitx} 
\usepackage{natbib}
\usepackage{color}
\usepackage{tabularray}
\usepackage[version=4]{mhchem}
\usepackage{amssymb}
\usepackage{amsmath}

\begin{document}

\title{Multiphase Reactive Transport in a Heterogeneous Flow Field: Channel Formation in Hydrocarbon-Bearing Carbonate Rock}

\author[label1,label2]{Qianqian Ma}
\author[label1,label2]{Zhuangzhuang Ma}
\author[label2]{Rukuan Chai}
\author[label1,label2]{Yanghua Wang}
\author[label2]{Martin J. Blunt}
\author[label2]{Branko Bijeljic\corref{cor1}}

\cortext[cor1]{Corresponding author.}
\ead{b.bijeljic@imperial.ac.uk}

\affiliation[label1]{%
  organization = {Resource Geophysics Academy, Imperial College London},
  addressline  = {London, SW7 2BP},
  country      = {United Kingdom}}

\affiliation[label2]{%
  organization = {Department of Earth Science and Engineering, Imperial College London},
  addressline  = {London, SW7 2AZ},
  country      = {United Kingdom}}

\begin{frontmatter}
\begin{abstract}
Carbon capture, utilisation and storage (CCUS) is a key approach for reducing anthropogenic CO\textsubscript{2} emissions. Currently the vast majority of CO$_2$ stored is injected into depleted hydrocarbon reservoirs. Reactive transport in hydrocarbon-bearing carbonate reservoirs is controlled not only by the balance between  reactant delivery and surface reaction, but also by the heterogeneous pore-scale flow field created by pore structure and fluid distribution. Here, we used time-resolved micro-CT imaging, pore-network analysis, and direct numerical simulation to investigate channelised dissolution during injection of CO\textsubscript{2}-saturated brine into oil-bearing Ketton limestone at 0.5~mL\,min$^{-1}$. The system remained in a high-$Pe$, low-$Da$ regime throughout the 180~min injection. Despite the advection-dominated flow regime, dissolution became strongly localised. Preferential flow pathways were already present originally because the heterogeneous pore structure and remaining-oil occupancy restricted brine flow to a subset of the connected pore space. Dissolution progressively amplified these pathways. The associated reduction in hydraulic resistance further focused flow, producing a continuous tortuous channel by 180~min. Meanwhile, the effective reaction rate was only $1.6\times10^{-5}$~mol\,m$^{-2}$\,s$^{-1}$, approximately one order of magnitude lower than the batch reaction rate, showing that rapid advective transport did not translate directly into rapid overall dissolution under multiphase conditions. These results demonstrate that channel formation arose from the coupling between strong reactant delivery and pore-scale flow-field heterogeneity, rather than from the bulk $Pe$--$Da$ regime alone. Accounting for flow-field heterogeneity is therefore important for predicting reactive transport and pore-structure evolution during CO\textsubscript{2} injection into hydrocarbon-bearing carbonate reservoirs.

\end{abstract}


\end{frontmatter}

\section{Introduction}
\label{sec1}
Geological storage of carbon dioxide (CO\textsubscript{2}) in depleted hydrocarbon reservoirs can use existing subsurface infrastructure while providing access to formations with well-characterised geological and petrophysical properties~\citep{zoback2023meeting, zoback2012earthquake}.  Currently more than 90 \% of the CO$_2$ injected worldwide is into oil fields \citep{gao2025london}. After injection, CO\textsubscript{2} dissolves in the resident brine and forms an acidic aqueous phase that reacts with carbonate minerals~\citep{andrew2013pore}. These reactions can alter pore geometry, reactive surface area, porosity and permeability, thereby influencing injectivity, fluid redistribution and the long-term evolution of the storage formation~\citep{al-khulaifiReservoirconditionPorescaleImaging2018, seyyedi2020pore, kampman2014fluid}. Depleted hydrocarbon reservoirs commonly retain hydrocarbons as disconnected or capillary-trapped ganglia after production, such that subsequent reactive transport occurs within a pore space reduced by the presence of the oil phase, rather than a fully brine-saturated rock~\citep{iglauer2012comparison, ma2026pore, pak2015droplet}. Understanding how remaining hydrocarbons affect dissolution and flow-path evolution is therefore necessary for predicting the behaviour of CO\textsubscript{2}-rich fluids in these formations.

In single-phase carbonate systems, dissolution patterns are governed by the competition between surface reaction, advection and molecular diffusion, commonly characterised using the P\'eclet and Damk\"ohler numbers \citep{golfier2002ability, fredd2000advances}. Depending on the transport regime and initial pore structure, dissolution may exhibit a uniform or compact pattern, or develop into preferential channels and wormholes \citep{golfier2002ability,menkeDynamicThreeDimensionalPoreScale2015,luquotExperimentalDeterminationPorosity2009}.
 Time-resolved imaging and numerical studies have shown that pore-scale heterogeneity plays a central role in selecting these pathways \citep{menkeDynamicReservoirconditionMicrotomography2017,noirielChangesReactiveSurface2009,al-khulaifiReservoirconditionPorescaleImaging2018,ma2026pore}.
P\'eclet and Damk\"ohler numbers characterize the competition among advective transport, diffusive transport and reaction, whereas pore-scale heterogeneity governs how the resulting reactant flux is distributed among available flow pathways. 

Remaining oil introduces an additional control because its impact on geometrical connectivity no longer guarantees a fully connected flow field. Oil occupying critical pores or throats can remove connections from the brine-conducting network, reduce brine–mineral contact and divert reactive fluid into a restricted set of accessible pathways~\citep{akindipePoreMatrixDissolution2022,ma2026pore, ma2026time}. CO\textsubscript{2} partitioning into the oil phase may cause swelling and further obstruct aqueous flow, whereas oil mobilisation or displacement can reopen pore throats and expose previously inaccessible mineral surfaces~\citep{riazi2011theoretical, alizadeh2014multi, ma2026pore, ma2026time}. The effect of remaining oil is therefore dynamic: it can suppress dissolution by limiting reactive access, but its subsequent movement can abruptly redistribute flow and intensify dissolution in newly opened regions. Ma et al.\citep{ma2026time} injected CO\textsubscript{2}-saturated brine into an oil-containing carbonate at 0.05 mL/min and found that mobilisation of remaining oil rapidly increased flow-field heterogeneity. Dissolution channels subsequently developed preferentially in pores from which oil had first been displaced, indicating that capillary opening controlled channel initiation under low-rate conditions.
Under high-flow rate conditions, strong advective penetration delivers a large reactant flux into the pore space, while the evolving oil–brine configuration redistributes this supply unevenly. Preferential pathways may receive more reactant, whereas oil redistribution and pore-throat enlargement modify local connectivity and redirect flow. Dissolution patterns therefore reflect the coupling between reactant delivery and multiphase flow heterogeneity. 
The objective of this study is to determine the impact of multiphase flow heterogeneity in a rock where the preferential flow paths already exist.

Here, we investigate the coupling between dissolution, oil swelling and mobilisation in a strongly preferential flow field during high-flow rate injection of CO\textsubscript{2}-saturated brine into a Ketton limestone sample containing oil.
To design strongly preferential flow field, before this stage, the sample underwent an earlier low-flow rate reactive injection, followed by removal of the remaining oil and a new drainage--imbibition cycle to establish the remaining oil configuration used in the present study. Time-resolved micro-computed tomography was used to visualize the evolution of the rock, brine and oil phases. Pore-network modelling characterised the pore--throat architecture and its progressive restructuring, while direct numerical simulation quantified the velocity field for the multiphase configuration. 
This framework is used to determine how high reactant flux interacts with a heterogeneous, dynamically evolving multiphase flow field to control the localisation and development of dissolution channels.

\section{Materials and Methods}
\subsection{Materials}
\label{subesc1}
A cylindrical sample of oolithic Ketton limestone was studied. The sample measured \SI{12}{\milli\meter} in length and \SI{6}{\milli\meter} in diameter. Calcite constituted \SI{99.1}{\percent} of the rock. The well-connected pore structure exhibited a bimodal pore-size distribution: the micro-pore-throat radius peaked below \SI{0.1}{\micro\meter}, whereas the macro-pore-throat radius peaked above \SI{10}{\micro\meter}. These two modes were separated by almost three orders of magnitude~\cite{patmonoaji2025differential}. The helium porosity was $0.239 \pm 0.008$~\cite{patmonoaji2025differential}.

Decane was selected as the oil phase \cite{akindipePoreMatrixDissolution2022}. The synthetic brine was prepared by dissolving 5~wt\% \ce{NaCl} and 1~wt\% \ce{KCl} in deionized water. To improve X-ray contrast during imaging, the solution was doped with 30~wt\% potassium iodide (\ce{KI})~\cite{lin2021drainage}. Under the experimental conditions of \SI{50}{\celsius} and \SI{8}{\mega\pascal}, the brine viscosity was \SI{0.82}{\milli\pascal\second}~\cite{patmonoaji2025differential}, whereas the corresponding decane viscosity was \SI{0.838}{\milli\pascal\second} (provided by PubChem, an open chemistry database).

\subsection{Experimental Methods and Image Acquisition}
\label{subsec2}

\textbf{i. Preparation of CO\textsubscript{2}-saturated brine.}
The brine (5~wt\% NaCl and 1~wt\% KCl, doped with 30~wt\% KI to enhance X-ray contrast) was pre-equilibrated with CO\textsubscript{2} in the injection pump at 8~MPa and \SI{50}{\celsius} for two weeks prior to injection.

\textbf{ii. Sample preparation and wettability alteration.}
The initially water-wet sample was enclosed in a Viton sleeve and mounted in a core holder surrounded by a heating jacket. The sample was first fully saturated with brine. Crude oil was then continuously injected at a constant flow rate of \SI{0.001}{\milli\liter\per\minute} for two weeks, while the pressure and temperature were maintained at \SI{8}{\mega\pascal} and \SI{60}{\celsius}, respectively. After one week, the flow direction was reversed, and crude-oil injection was continued at the same rate for a further week. This procedure, called ageing, was conducted as a separate batch experiment before micro-CT imaging to alter the initially water-wet surfaces and establish mixed-wet conditions. Following ageing, the sample was sequentially flushed with decalin and decane. The system was then depressurized and disassembled to allow the sample to be transferred and installed in the flow apparatus mounted on the micro-CT scanner.

\begin{figure}[H]
  \centering
  \includegraphics[width=1\textwidth]{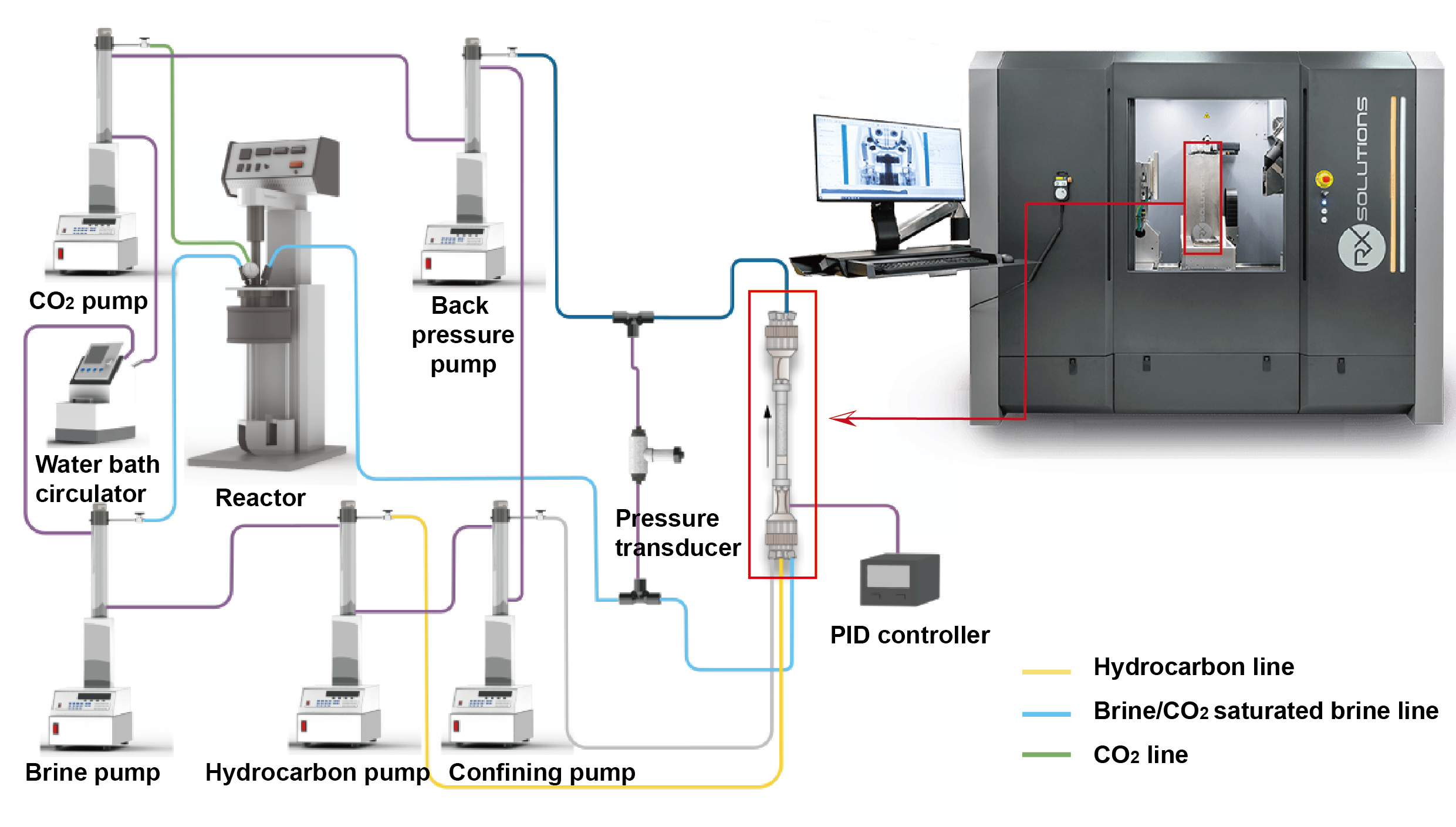}
  \caption{The experimental apparatus including the flow loop and micro-CT scanner.}
  \label{fig:apparatus}
\end{figure}

\textbf{iii. System assembly.}
The aged sample was reassembled in the core holder and installed in a CT scanner (RX Solution Easy Tom L X-ray Microscope). The core holder was then connected to the fluid injection and production lines and to the pressure transducers, as shown in Figure~\ref{fig:apparatus}. A confining pressure of 10~MPa and a back pressure of 8~MPa were applied, while the sample temperature was maintained at \SI{50}{\celsius} throughout the experiment. Fluids were injected through the bottom inlet.

\textbf{iv. Brine injection.}
The doped brine was injected at 0.05~mL\,min$^{-1}$ for 100 pore volumes (PVs), after which a micro-CT scan was acquired.

\textbf{v. Establishment of initial oil saturation (\(S_{\mathrm{oi}}\)).}
Decane was injected at 0.05~mL\,min$^{-1}$ for 100 PVs to establish \(S_{\mathrm{oi}}\), after which a micro-CT scan was acquired.

\textbf{vi. Establishment of remaining oil saturation (\(S_{\mathrm{or}}\)).}
Brine was re-injected at 0.05~mL\,min$^{-1}$ for 170 PVs, giving \(S_{\mathrm{or}} = 25.12\%\); the remaining oil configuration was imaged before the reactive stage commenced.

\textbf{vii. Low-flow rate reactive injection.}
CO\textsubscript{2}-saturated brine was injected at 0.05~mL,min$^{-1}$, during which the displacement was monitored by a series of time-lapse micro-CT scans. At the end of this stage, non-reactive brine was injected to displace all remaining oil from the sample. This stage is not the focus of the present work.

\textbf{viii. Re-establishment of remaining oil saturation.}
Before decane injection, a micro-CT scan of the brine-saturated sample was acquired to characterize the pore structure after dissolution. Decane injection followed by brine injection, both conducted at 0.05~mL\,min$^{-1}$, then re-established a remaining oil configuration within the partially dissolved pore space. The resulting fluid distribution was imaged and used as the baseline for the second reactive injection, corresponding to \SI{0}{\minute}. The oil--brine configuration in this image indicated that the core had become water-wet after the preceding dissolution.

\textbf{ix. High-flow rate reactive injection.}
CO\textsubscript{2}-saturated brine was re-injected at
0.5~mL\,min$^{-1}$ for 180~min; this stage is the focus of the
present study. During reactive injection, six sequential micro-CT scans were acquired at \SI{30}{\minute} intervals, corresponding to \(t = 30\), 60, 90, 120, 150, and \SI{180}{\minute}.
Each scan was acquired using 1504 projections. The X-ray source was operated at a tube voltage of \SI{146}{\kilo\volt} and a tube current of \SI{175}{\micro\ampere}. The projection images
contained $1944 \times 2456$ pixels, and the reconstructed images had an isotropic voxel size of \SI{5.2}{\micro\meter}. The resulting field of view was approximately \SI{10.11}{\milli\meter}~\texttimes~ \SI{10.11}{\milli\meter}~\texttimes~ \SI{12.77}{\milli\meter}, encompassing the full \SI{6}{\milli\meter}-diameter and \SI{12}{\milli\meter}-long core.
Each scan required approximately \SI{30}{\minute}. Scan times denote the cumulative acquisition time relative to the remaining-oil scan at \(t=0\); idle periods between consecutive acquisitions were excluded.

\begin{table}[!htbp]
\centering
\caption{Time-resolved micro-CT imaging protocol. All scans were acquired
             during continuous injection of CO$_2$-saturated brine at
             $0.5\;\mathrm{mL\;min^{-1}}$. Cumulative pore volumes (PV) are
             referenced to the end of each acquisition window.}
\sisetup{table-format=4.0}
\begin{tabular}{
  l 
  c 
  S[table-format=4.0]
}
\toprule
\textbf{Scan} & \textbf{Time interval (Start--End)} & \textbf{Pore volumes} \\
& {(\si{\minute})} & {} \\
\midrule
0 min scan & \multicolumn{2}{c}{Remaining oil saturation scan} \\
\midrule
\SI{30}{\minute} scan & t=0 to t=30   & {230} \\
\SI{60}{\minute} scan & t=30 to t=60 & {450} \\
\SI{90}{\minute} scan & t=60 to t=90 & {660} \\
\SI{120}{\minute} scan & t=90 to t=120 & {861} \\
\SI{150}{\minute} scan & t=120 to t=150 & {1052} \\
\SI{180}{\minute} scan & t=150 to t=180 & {1240} \\
\bottomrule
\end{tabular}
\label{tab:pv_comparison}
\end{table}

\subsection{Image Processing and Analysis}

Three-dimensional tomograms were reconstructed using RX-Solutions Reconstructor software.
Image processing was conducted in commercial image analysis software (Avizo) using a standardized workflow to ensure cross-scan comparability. Initially, images were denoised using a combination of non-local means and anisotropic diffusion filters. This was followed by intensity normalization to standardize greyscale ranges. All scans were co-registered to a common reference. This systematic approach guaranteed consistency across all imaging stages, facilitating accurate segmentation and quantitative analysis. Phase segmentation was performed to distinguish three phases: rock matrix, brine, and oil.

\textbf{Pre-reaction dataset.}
The reference segmentations were generated using a hybrid workflow combining differential imaging~\cite{Gao2017,chai2025pore,chai2022formation}, interactive threshold, and watershed-based segmentation~\cite{Lin2016}. Specifically, (i) the rock mask was extracted from the brine reference scan; (ii) oil was segmented directly from the greyscale images without differential imaging; and (iii) brine was identified by subtracting the dry scan from the brine- or oil-saturated scans, thereby identifying pores occupied by the aqueous phase. 

\textbf{Model training and fine-tuning.}
Time-series segmentation was performed using PoreTimeNet~\cite{ma2026spatiotemporal} , a spatiotemporal deep learning framework. The initial model was trained on a separate labelled Ketton dataset~\cite{ma2026time}. The pretrained model was subsequently fine-tuned on the reactive Ketton dataset using the reference segmentations as sparse target-domain supervision. The fine-tuned model was then applied to the complete time series to generate the phase segmentations.

\textbf{Time-series inference.}
The fine-tuned model was applied to the other normalized greyscale volumes as a time series to generate the final rock, oil, and brine segmentations.

\subsection{Pore-Scale Simulation}
\label{sec:2.5}
Following image segmentation, the lower region of the segmented micro-CT volume ($5.72 \times 5.72 \times 5.2 \ \mathrm{mm}^3$) was selected for pore-scale flow simulation in the brine phase. The modelling approach follows the method developed by  Bijeljic et al. \cite{bijeljicInsightsNonFickianSolute2013} and Raeini et al. \cite{RAEINI20125653}, implemented within the \texttt{OpenFOAM} framework. The solver employs the finite volume method to simultaneously solve the continuity and steady-state incompressible Navier–Stokes equations:
\begin{align}
\nabla \cdot \mathbf{u} &= 0 \\
\rho \left( \frac{\partial \mathbf{u}}{\partial t} + \mathbf{u} \cdot \nabla \mathbf{u} \right) &= -\nabla p + \mu \nabla^2 \mathbf{u}
\label{NS}
\end{align}
where $p$ is the pressure ($\mathrm{Pa}$), and $u$ is the velocity ($\mathrm{m\,s^{-1}}$), both obtained for each voxel of the image; $\mu$ is the fluid (brine) viscosity ($\mathrm{Pa\,s}$); $\rho$ is the fluid density ($\mathrm{kg\,m^{-3}}$). The flow rate ($\mathrm{m^3\,s^{-1}}$) is calculated as $Q = \int u_x \, dA_x$, where $A_x$ is the cross-sectional area of the image ($\mathrm{m^2}$) and $u_x$ is the velocity in the direction of overall flow ($\mathrm{m\,s^{-1}}$). The Darcy velocity ($\mathrm{m\,s^{-1}}$) is then calculated as $q = \frac{Q}{L_y L_z}$, where $L_y$ and $L_z$ are the lengths of the image ($\mathrm{m}$).
A fixed 1 Pa pressure drop was applied along the flow direction, while all solid boundaries were treated as no-slip walls. In Section 3.3, the reported streamlines specifically correspond to the flow of brine in the presence of remaining oil. These values were determined by simulating the flow field within the connected, brine-saturated pore network, where the remaining oil was assumed to remain immobile. 

Additionally, the network extraction code based on the maximal ball algorithm \citep{PhysRevE.80.036307} was used to quantify the geometric and topological properties of the pore space. In the maximal ball algorithm, spheres are generated in the void space of the segmented images to determine the positions and diameters of pores; throats are the restrictions between pores. In this way, the pore space is topologically represented as a network of pores connected by narrow throats. 

\subsection{Dimensionless numbers and reaction rates}
In this study, the dimensionless number, Péclet (Pe) and Damköhler (Da), are used to characterize and quantify the reactive transport.
The Péclet number quantifies the relative efficiency of solute mass transfer through advection compared to diffusion \cite{peclet1827traite}:
\begin{align}
   \text{Pe} = \frac{\text{advective transport rate}}{\text{diffusive transport rate}} = \frac{u_{\text{av}} L_c}{D_m}
   \label{Pe}
\end{align}
where $D_m$ is molecular diffusion coefficient in brine ($\mathrm{m^2\,s^{-1}}$), $u_{\text{av}}$ is the average pore velocity, which is the Darcy velocity from the experiment divided by the product of porosity and brine saturation, while $L_c$ is characteristic length ($m$), calculated by \cite{mostaghimi2012simulation}:
\begin{align}
    L_c = \frac{\pi}{S}
\end{align}
where the specific surface area S (m$^{-1}$), is the image surface area per unit volume at the beginning of the time period, calculated by $V_B/As$, where $V_B$ is bulk volume, and $As$ is the surface area from image analysis.

The Damköhler number quantifies the ratio between the timescales of a chemical reaction and the mass transfer \cite{lasaga1984chemical}:
\begin{align}
    \text{Da} = \frac{\text{reaction rate}}{\text{advective transport rate}} = \frac{L_c}{u_{\text{av}}} k
    \label{12}
\end{align}
where $k$ is the chemical reaction rate constant (s$^{-1}$), calculated by: $ k = \frac{\pi r}{nL}$.
r is the mineral reaction rate ($\mathrm{mol\,m^{-2}\,s^{-1}}$), L is the sample length ($m$), and $n$ is calculated by \cite{menkeDynamicThreeDimensionalPoreScale2015}:
\begin{align}
    n = \frac{\rho_{\text{mineral}} f_{\text{mineral}}}{M_{\text{mineral}}}
\end{align}
where $\rho_{\text{}}$ is the mineral density, $M_{\text{mineral}}$ is their molecular mass. Therefore, equation \ref{12} can be rewritten as follows\cite{al-khulaifiReservoirconditionPorescaleImaging2018}:
\begin{align}
    \text{Da} = \frac{\pi r_{\text{mineral}}}{u_{\text{av}} n}
    \label{Da}
\end{align}
where $r_{\text{}}$ is the non-transport limited reaction rate. 

The effective reaction rate ($r_{\text{eff}}$) of mineral is determined as \cite{al-khulaifiReactionRatesChemically2017a}:
\begin{align}
    r_\text{eff} = \frac{\rho_\text{mineral} (1 - \phi_\text{unresolved}) \Delta \phi_\text{CT}}{M_\text{mineral} S \Delta t}
    \label{eq:reaction rate}
\end{align}
where $\Delta t$ is the time between scans (s), $\Delta \phi_\text{CT}$ is the corresponding change in porosity and $S$ is the image surface area per unit volume (\(\text{m}^{-1}\)). We observe almost no change in greyscale values for solid voxels containing sub-resolution features, indicating no measurable change in porosity in these regions. The unresolved porosity ($\phi_\text{unresolved}$) is calculated based on data from a sister sample drilled from the same block \cite{patmonoaji2025differential, ma2026pore}.

\section{Results and Discussion}

Sections~\ref{subsec:3.1}--\ref{subsec:3.4} examine how dissolution, oil swelling, and mobilisation interact to control channel development during reactive injection. We first characterise the time-resolved evolution of the pore structure and oil distribution (Sections~\ref{subsec:3.1}), then quantify the accompanying pore-network changes (Sections~\ref{subsec:3.2}) and preferential flow pathways (Sections~\ref{subsec:3.3}). Finally, these observations are integrated with the $Pe$--$Da$ regime and effective reaction rate to identify how reactant delivery couples with the evolving multiphase flow field to produce channelised dissolution (Sections~\ref{subsec:3.4}).

\subsection{Time-Resolved Development of Channelised Dissolution}
\label{subsec:3.1}

\begin{figure}[H]
    \centering
    \includegraphics[width=1\textwidth]{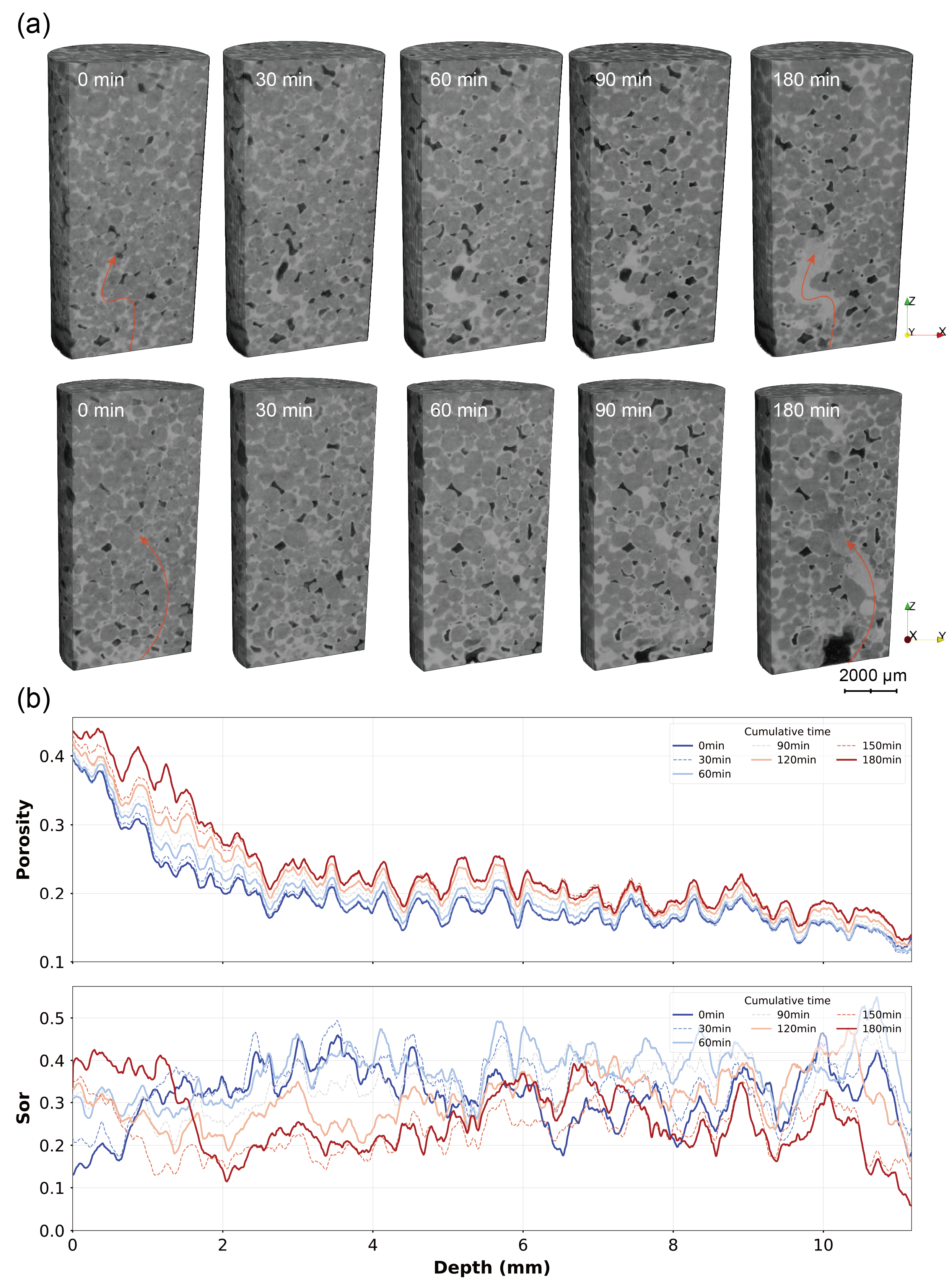}
    \caption{Time-resolved evolution of the pore structure and remaining-oil distribution during the second injection of CO$_2$-saturated brine. (a) Three-dimensional greyscale renderings at 0, 30, 60, 90, and 180~min from two viewing orientations. Gray represents solid, black represents oil, and white represents brine. Orange arrows indicate the preferential pathway along which the dissolution channel subsequently developed. (b) Axial profiles of porosity (top) and remaining oil saturation, $S_{\mathrm{or}}$ (bottom), from 0 to 180~min. Scale bar: $2000~\mu\mathrm{m}$.}
    \label{fig:CT}
\end{figure}

Figure~\ref{fig:CT} shows the time-resolved evolution of the pore structure and remaining-oil distribution during the high-flow rate injection of CO$_2$-saturated brine. Dissolution did not produce a new channel at a random location. Instead, progressive pore enlargement followed a curved pathway that could already be identified from the initial arrangement of connected, relatively oil-free pores. The two viewing orientations show consistent spatial evolution, confirming that the developing channel was a three-dimensional feature within the sample.

At 0~min, no continuous dissolution channel was evident, although a sequence of connected pores along the trajectory indicated by the orange arrows provided a potential pathway (Figure~\ref{fig:CT}a). During the first 30~min, structural changes were limited and mainly involved slight enlargement of pores and throats along this pathway. By 60~min, these features had widened further and become increasingly connected. Between 60 and 90~min, neighbouring pores progressively enlarged and coalesced, producing a distinct high-porosity pathway. Continued dissolution up to 180~min further widened this region, ultimately forming a continuous channel with an irregular width and tortuous morphology, while dissolution remained less pronounced in the surrounding pore space.

The axial porosity profiles provide complementary evidence for this structural evolution (Figure~\ref{fig:CT}b). Porosity increased progressively over most of the imaged length, indicating that dissolution was not confined to a narrow region along the sample length. Nevertheless, the three-dimensional images show that pore enlargement was spatially concentrated within the developing channel. 

In contrast, remaining oil saturation exhibited a strongly heterogeneous and non-monotonic evolution along the sample. Local increases in $S_{\mathrm{or}}$ were consistent with CO\textsubscript{2} partitioning into decane and the associated oil swelling, whereas subsequent decreases reflected oil mobilisation and redistribution. Under low-flow conditions, Ma et al.\ showed that oil swelling can obstruct conductive pores and throats, restrict acid accessibility, and suppress the effective dissolution rate~\citep{ma2026time}. In the present high-flow rate experiment, rapid renewal of CO\textsubscript{2}-saturated brine at oil--brine interfaces promoted CO\textsubscript{2} transfer into the oil phase and pronounced swelling. Meanwhile, the larger viscous pressure gradients could mobilise oil ganglia close to their capillary trapping thresholds. The observed $S_{\mathrm{or}}$ evolution therefore reflected competition between swelling-induced obstruction and viscous mobilisation, maintaining a dynamically heterogeneous flow field during dissolution.
\newgeometry{top=2.5cm,bottom=2.5cm}
\restoregeometry
From 60~min onwards, relatively large low-attenuation regions also appeared near the sample inlet and within the developing preferential channel (Figure~1a). These features are interpreted as the possible formation of a free CO\textsubscript{2}-rich phase. Strong dissolution near the inlet substantially enlarged the local pore space, while the high flow rate generated appreciable viscous pressure gradients. A sufficient local pressure decrease could shift the CO\textsubscript{2}--brine system across its phase-equilibrium condition and induce CO\textsubscript{2} exsolution. Similar low-attenuation features within the preferential pathway suggest that this process may also have occurred locally inside the channel. 

These observations show that broad reactant penetration coexisted with strongly localised structural alteration. The relationship between this initial multiphase heterogeneity, preferential flow, and subsequent channel development is examined quantitatively using pore-network analysis and direct numerical simulations in the following sections.

\subsection{Pore-Network Structure and its Evolution during Dissolution}
\label{subsec:3.2}
\begin{figure}[H] 
\centering 
\includegraphics[width=1\textwidth]{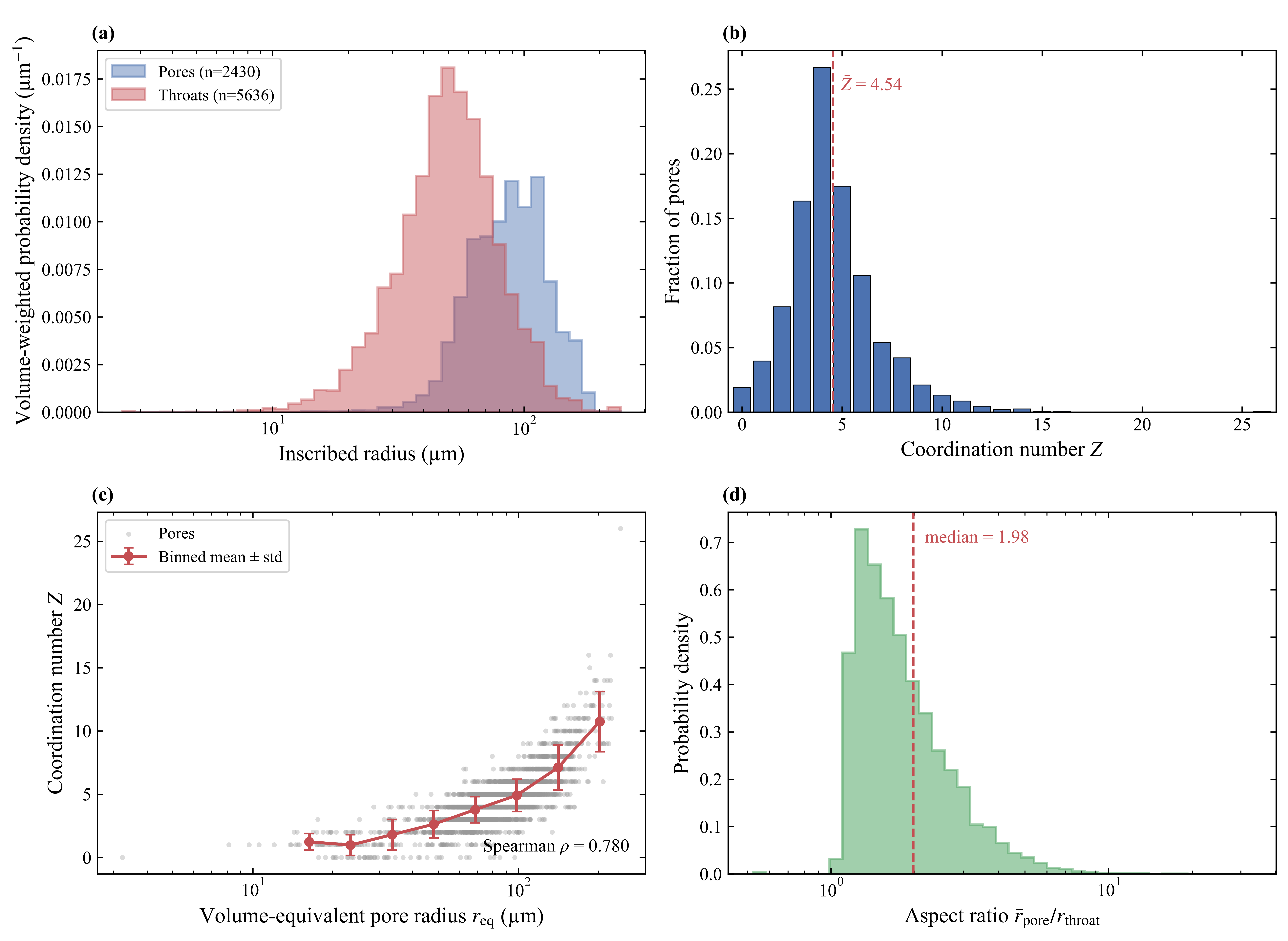} 
\caption{Pore-network geometry and topology at 0~min of high-flow rate injection: (a)~volume-weighted probability density of inscribed pore and throat radii; (b)~coordination-number distribution, with the mean $\bar{Z} = 4.54$ marked; (c)~coordination number $Z$ against volume-equivalent pore radius $r_{\mathrm{eq}} = (3V/4\pi)^{1/3}$, showing individual pores (grey) and the binned mean $\pm$ one standard deviation (Spearman $\rho = 0.78$); (d)~distribution of the pore--throat aspect ratio $\bar{r}_{\mathrm{pore}}/r_{\mathrm{throat}}$ over internal throats, where $\bar{r}_{\mathrm{pore}}$ is the mean inscribed radius of the two adjacent pores (median 1,98).}
\label{fig:pnm0} 
\end{figure}

At the start of high-flow rate reactive injection, the extracted pore network contained 2430 pores and 5636 throats (Figure~\ref{fig:pnm0}). The pore-size distribution was shifted towards larger radii relative to the throat-size distribution (Figure~\ref{fig:pnm0}a). The network had a mean coordination number of 4.54, with most pores having coordination numbers between approximately 2 and 7 and a smaller population extending to higher values (Figure~\ref{fig:pnm0}b). Pore size was positively correlated with coordination number (Spearman $\rho = 0.78$), indicating that larger pores generally had more connections (Figure~\ref{fig:pnm0}c). The pore--throat aspect-ratio distribution was right-skewed, with a median of 1.98, showing that the degree of pore--throat constriction varied across the network (Figure~\ref{fig:pnm0}d).

The influence of the low-flow rate reactive injection becomes evident when this structure is compared with that before reaction (Table~\ref{tab:pnm_before_after}). The changes were substantially greater at the extremes of the distributions than in their bulk properties. The maximum throat radius increased by 79\%, compared with 17\% at the 95th percentile, while the maximum pore radius increased by 41\%, compared with 11\% at the 95th percentile. Similarly, the maximum coordination number increased by 86\%, whereas the mean increased by only 11\%. The fraction of highly connected pores increased by 54\%, while the fraction of isolated pores increased by 46\%. These disproportionate changes indicate that the low-flow rate dissolution did not modify the pore network uniformly, but increased the contrast between highly enlarged and connected regions and the remainder of the network. The high-flow rate reactive injection therefore started from a pore structure whose heterogeneity had already been enhanced by its previous dissolution history.

\begin{figure}[H]
    \centering
    \includegraphics[width=1\textwidth]{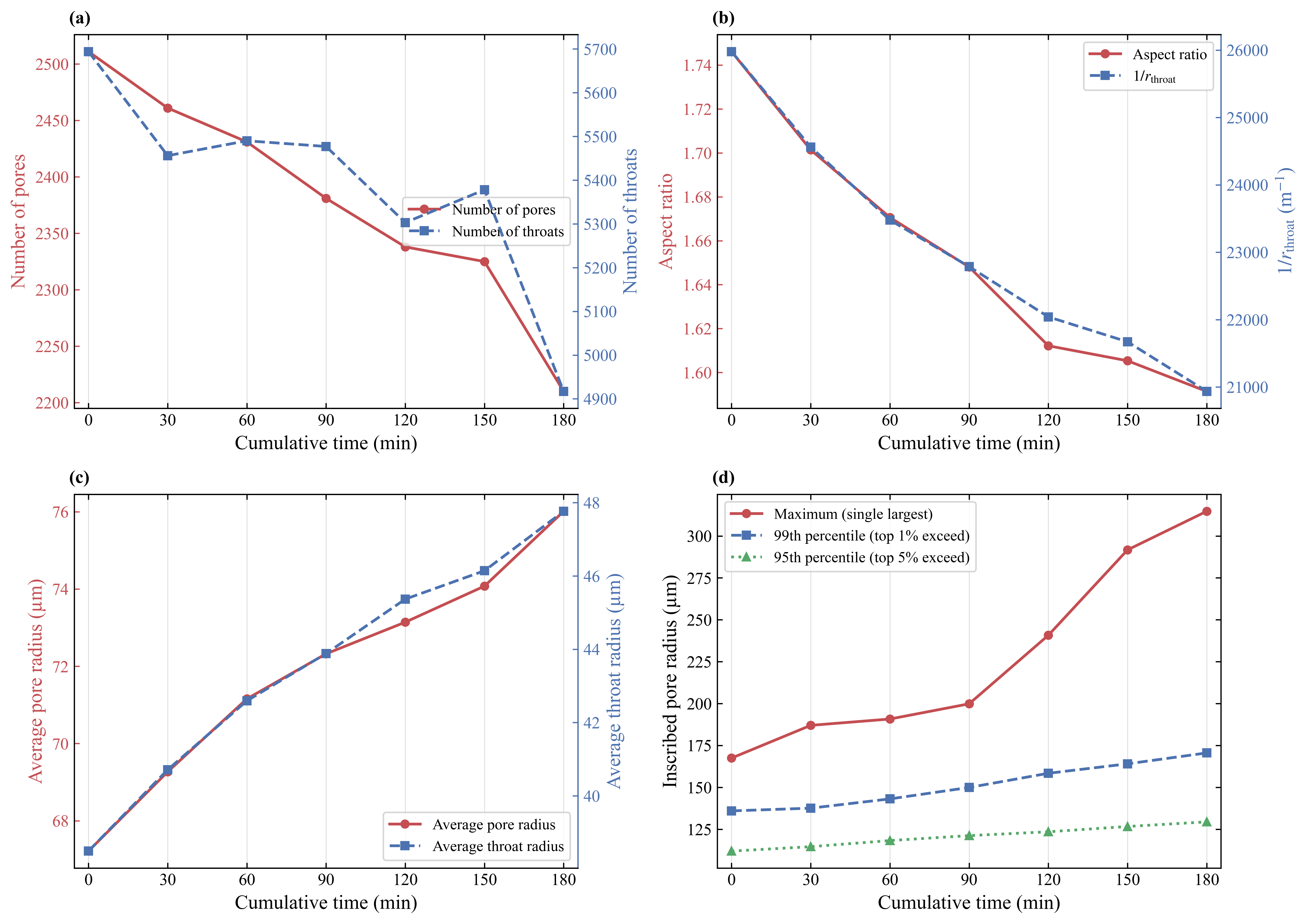}
    \caption{Pore-network geometry and topology during the high-flow rate reactive
    injection of CO$_2$-saturated brine (0.5~mL/min), extracted with
    \textit{pnextract} from the time-series micro-CT images. Evolution over 0--180~min:
    (a)~number of pores and throats;
    (b)~mean pore--throat aspect ratio and inverse mean throat radius
    $1/r_{\mathrm{throat}}$, a proxy for the capillary entry pressure;
    (c)~mean inscribed pore and throat radii;
    (d)~largest inscribed pore radius, shown as the single maximum together
    with the 99th and 95th percentiles: the maximum grows far faster than
    the percentiles, indicating that dissolution is concentrated on the
    largest pores.}
    \label{fig:pnm}
\end{figure}

During CO\textsubscript{2}-saturated brine injection, both pores and throats enlarged progressively (Figure~\ref{fig:pnm}a--d). The mean pore radius increased by approximately 13\%, whereas the mean throat radius increased by approximately 24\%. Consequently, the mean pore--throat aspect ratio decreased from 1.75 to 1.59 and $1/r_{\mathrm{throat}}$ declined continuously, indicating preferential widening of the constrictions controlling local hydraulic and capillary resistance. Over the same period, the numbers of identified pores and throats decreased by approximately 12\% and 14\%, respectively. These changes are consistent with progressive pore coalescence and network coarsening as dissolution widened connecting throats and merged neighbouring pore bodies.

The maximum pore radius and the 95th and 99th percentiles were compared to assess whether pore enlargement became concentrated within the upper tail of the pore-size distribution (Figure~\ref{fig:pnm}d). The maximum pore radius increased from approximately 167 to 315~$\mu$m, nearly doubling over 180~min, whereas the 95th and 99th percentiles increased by only approximately 16\% and 25\%, respectively. The divergence became particularly pronounced after 90~min, coincident with the emergence of the channel observed in the micro-CT images. Thus, dissolution involved both widespread pore-throat enlargement and disproportionately rapid growth within a small subset of pores, producing the structural signature of a locally coarsened dissolution pathway. The influence of this evolving structure and remaining-oil occupancy on the preferential flow field is examined directly using DNS in the following section.

\begin{table}[htbp]
  \centering
  \caption{Pore-network metrics of the bottom part of the core before and
    after the low-flow rate reactive injection.}
  \label{tab:pnm_before_after}
  \begin{tabular}{lccc}
    \hline
    Metric & Before Reaction & After Low-Flow rate Injection & Change \\
    \hline
    Maximum coordination number $Z_{\max}$        & 14    & 26    & $+86\%$ \\
    Maximum throat inscribed radius (\si{\micro\metre})  & 133.8 & 239.8 & $+79\%$ \\
    Fraction of highly connected pores ($Z \geq 10$) & 2.12\% & 3.25\% & $+54\%$ \\
    Fraction of isolated pores ($Z = 0$)          & 1.29\% & 1.89\% & $+46\%$ \\
    Maximum pore inscribed radius (\si{\micro\metre})    & 171.1 & 241.7 & $+41\%$ \\
    P99 throat inscribed radius (\si{\micro\metre})      & 91.3  & 117.1 & $+28\%$ \\
    P99 pore inscribed radius (\si{\micro\metre})        & 135.2 & 159.8 & $+18\%$ \\
    P95 throat inscribed radius (\si{\micro\metre})      & 70.6  & 82.5  & $+17\%$ \\
    Mean coordination number $\bar{Z}$            & 4.10  & 4.54  & $+11\%$ \\
    P95 pore inscribed radius (\si{\micro\metre})        & 110.8 & 123.3 & $+11\%$ \\
    \hline
  \end{tabular}
\end{table}

\subsection {Evolution of Preferential Flow Pathways}
\label{subsec:3.3}

Direct numerical simulations were used to quantify how remaining-oil occupancy and subsequent dissolution modified the flow field. For the without oil reference case, the entire resolved pore space was available to brine flow, whereas in the with oil cases the oil phase was treated as impermeable. This comparison therefore isolates the influence of phase occupancy on the distribution of flow field.

\begin{figure}[H]
    \centering
    \includegraphics[width=\textwidth]{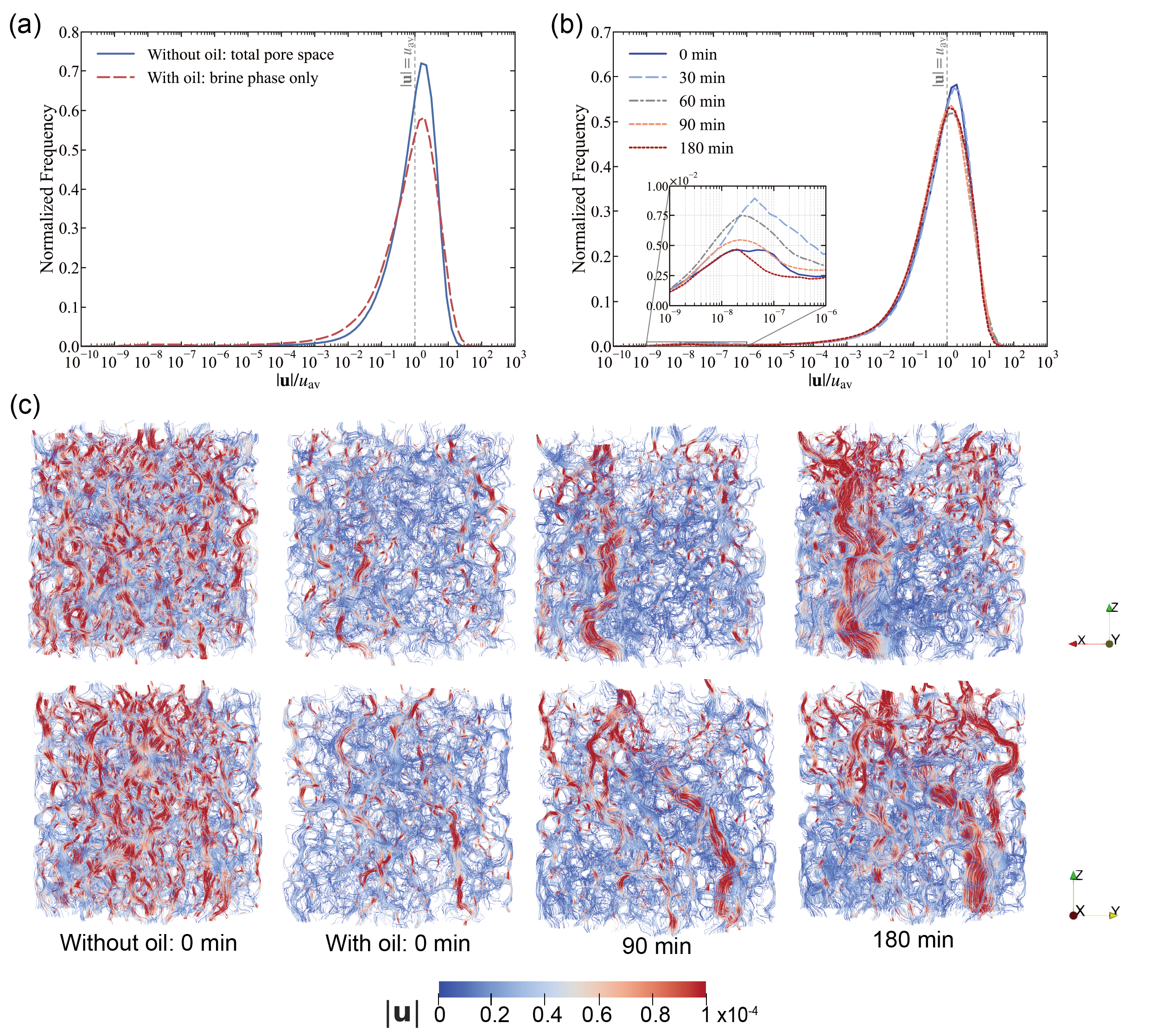}
    \caption{
    Evolution of the pore-scale flow field during high-flow rate reactive injection.
    (a) Normalised probability distributions of the velocity magnitude at 0~min 
    for the oil-free total pore space and the brine-accessible pore space containing remaining oil.
    The velocity magnitude, $|\mathbf{u}|$, is normalised by the average velocity, $u_{\mathrm{av}}$.
    (b) Evolution of the normalised velocity distribution at 0, 30, 60, and 90~min; 
    the inset highlights changes in the low-velocity region.
    (c) Streamline visualisations of the velocity field in two orthogonal views.
    Comparison of the oil-free and oil-containing cases at 0~min shows that remaining-oil occupancy 
    concentrates flow into a smaller subset of connected pathways, while the 60 and 90~min fields 
    show progressive focusing and strengthening of preferential flow pathways during dissolution.
    Streamlines are coloured by velocity magnitude.
    }
    \label{fig:dns}
\end{figure}

At 0~min, the presence of remaining oil broadened the velocity distribution relative to the oil-free case (Figure~\ref{fig:dns}a), indicating a more heterogeneous flow field. By occupying individual pores and throats, the oil phase restricted parts of the brine-conducting network and diverted flow into a smaller subset of connected pathways. Consistent with this redistribution, the streamline visualisations show a more broadly distributed flow field in the oil-free pore space, whereas remaining-oil occupancy concentrated the flow into a smaller number of distinct high-velocity pathways (Figure~\ref{fig:dns}c). Preferential flow pathways were therefore already established before the high-flow rate reactive injection began. The initial flow field reflected the combined influence of the dissolution-modified pore structure described in Section~\ref{subsec:3.2} and the additional flow heterogeneity imposed by remaining-oil occupancy.

The velocity distributions became broader during CO\textsubscript{2}-saturated brine injection (Figure~\ref{fig:dns}b), indicating increasing separation between fast-flowing pathways and weakly supplied regions. The streamline visualisations provide the corresponding spatial evidence (Figure~\ref{fig:dns}c). At 0~min, several preferential flow pathways were already visible. By 30~min, flow became more concentrated within a subset of these pathways, and by 60~min, coherent high-velocity pathways had become markedly more pronounced. Their locations corresponded to the regions of preferential pore enlargement observed in the time-resolved micro-CT images (Figure~\ref{fig:CT}a).

These results show that dissolution did not generate preferential flow from an initially uniform flow field. Instead, reactive injection acted on a pre-existing heterogeneous multiphase flow field. Preferentially supplied regions underwent greater pore and throat enlargement, while the resulting reduction in hydraulic resistance further concentrated flow into these regions. The initial preferential pathways were therefore progressively amplified as dissolution, pore-network restructuring and phase occupancy evolved. This evolving flow heterogeneity provides the spatial framework for interpreting the coupling between reactant delivery and dissolution in the following section.

\subsection{Coupling between Reactant Delivery and Multiphase Flow Heterogeneity}
\label{subsec:3.4}
\begin{figure}[H]
    \centering
    \includegraphics[width=0.95\textwidth]{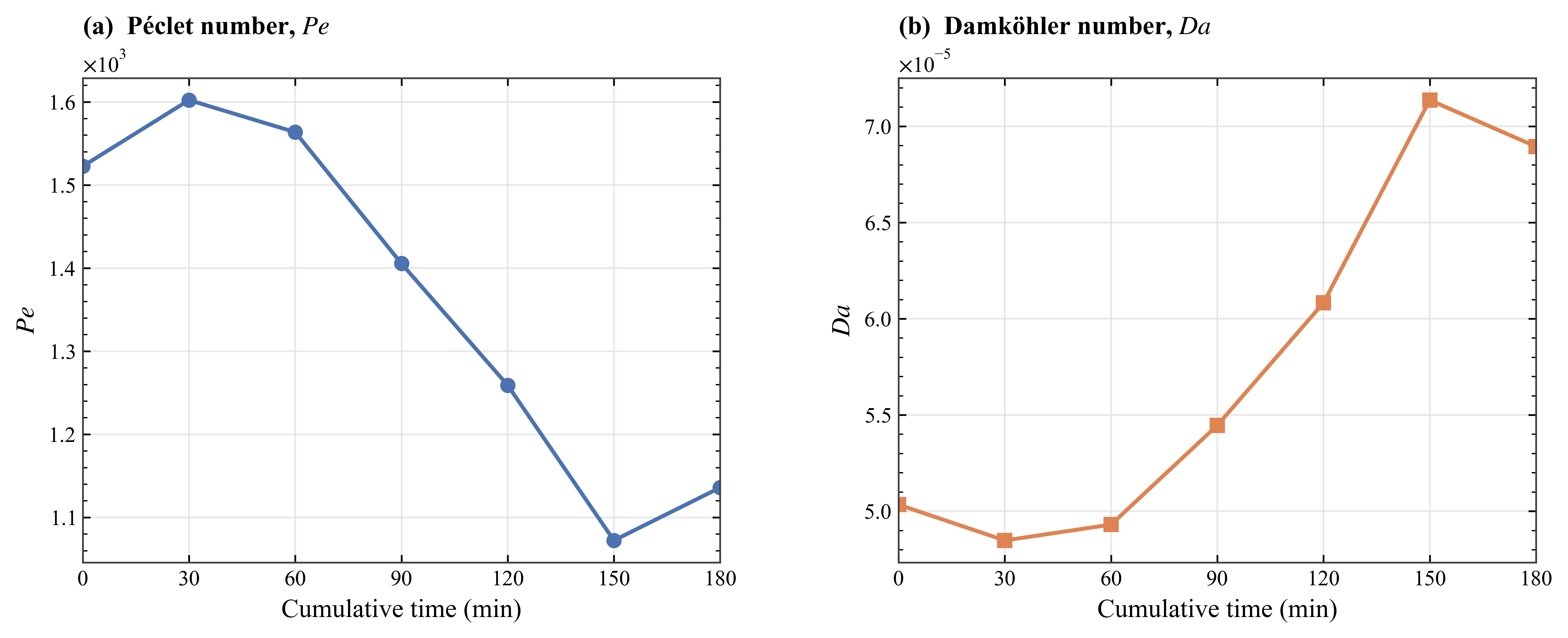}
    \caption{
    Evolution of the dimensionless transport--reaction numbers during the high-flow rate injection of CO$_2$-saturated brine.
    (a) Péclet number, $Pe$;
    (b) Damköhler number, $Da$;
    The system remained within a high-$Pe$, low-$Da$ regime throughout the injection.
    }
    \label{fig:PeDa}
\end{figure}

The high-flow rate reactive injection remained within a high-$Pe$, low-$Da$ regime throughout the experiment (Figure~\ref{fig:PeDa}). The Péclet number remained of order $10^{3}$, ranging from approximately 1070 to 1600, whereas the Damköhler number remained of order $10^{-5}$, between approximately $4.8\times10^{-5}$ and $7.1\times10^{-5}$. These values indicate that advection was rapid relative to molecular diffusion and mineral reaction, enabling reactive brine to penetrate deeply through the hydraulically accessible pore space before substantial local consumption. Although both $Pe$ and $Da$ evolved as dissolution modified the pore structure, the system remained within the same advection-dominated regime.

Despite this strong advective transport, the overall effective reaction rate was only $1.6\times10^{-5}$~mol\,m$^{-2}$\,s$^{-1}$. This was approximately one-third of the $5.0\times10^{-5}$~mol\,m$^{-2}$\,s$^{-1}$ reported for single-phase reactive flow through Ketton limestone at $Pe=933$~\citep{menkeDynamicThreeDimensionalPoreScale2015}, and more than one order of magnitude below the batch reaction rate of $6.9\times10^{-4}$~mol\,m$^{-2}$\,s$^{-1}$~\citep{peng2015kinetics}. Thus, a higher advective transport rate did not translate directly into a higher effective dissolution rate under multiphase conditions.

This apparent discrepancy can be explained by the spatial distribution of reactant delivery. The high-$Pe$, low-$Da$ regime enabled rapid transport of reactive brine, but the dissolution-modified pore structure and remaining-oil occupancy produced a heterogeneous flow field that concentrated this supply within a restricted set of preferential pathways. Regions receiving greater reactant flux dissolved more rapidly, which enlarged pores and throats and further reduced their hydraulic resistance. Meanwhile, the surrounding pore space was increasingly bypassed. Channelling therefore resulted from the coupling between strong reactant delivery and its heterogeneous spatial distribution, rather than from the bulk transport--reaction regime alone.

\begin{table}[htbp]
    \centering
    \caption{Comparison of the reaction rate measured in our multiphase
             experiment with single-phase results from the literature and
             the batch reaction rate. P\'eclet numbers are evaluated before
             the onset of reaction.}
    \label{tab:reaction_rates}
    \begin{tabular}{l S[table-format=1.1e-1] S[table-format=4.0]}
        \toprule
        Condition & {Reaction rate (\unit{\mol\per\square\metre\per\second})} & {\textit{Pe}} \\
        \midrule
        Ketton limestone, multiphase (this work)                                & 1.6e-5 & 1523 \\
        Ketton limestone, single phase \cite{menkeDynamicThreeDimensionalPoreScale2015} & 5.0e-5 &  933 \\
        Batch reaction \cite{peng2015kinetics}                                  & 6.9e-4 & {--}  \\
        \bottomrule
    \end{tabular}
\end{table}

\section{Conclusions}

We combined time-resolved micro-CT imaging, pore-network analysis, and direct numerical simulation to examine channel formation during a 0.5~mL/min injection of CO\textsubscript{2}-saturated brine through oil-bearing Ketton limestone over 180~min. The central finding was that dissolution did not create a channel at a random location: it amplified a pre-existing preferential flow structure, with a distinct high-porosity pathway emerging and developing into a continuous, tortuous channel.

The high-flow rate reactive injection acted on an already heterogeneous pore structure, with remaining-oil occupancy further concentrating brine flow into preferential pathways that were present initially. Dissolution progressively amplified these pathways: the mean pore and throat radii increased by approximately 13\% and 24\%, while the maximum pore radius increased from approximately 167 to 315~$\mu$m, far exceeding the approximately 16\% and 25\% increases at the 95th and 99th percentiles. The resulting reduction in local hydraulic resistance further focused flow, indicating that channel formation arose from amplification of pre-existing preferential pathways rather than from an initially uniform flow field.

The bulk transport--reaction regime alone did not determine either channel localisation or the effective dissolution rate. The experiment remained at high $Pe$ and low $Da$, with $Pe$ ranging from approximately 1070 to 1600 and $Da$ from approximately $4.8\times10^{-5}$ to $7.1\times10^{-5}$, conditions that enabled deep advective penetration through hydraulically accessible pores. Nevertheless, the effective reaction rate was only $1.6\times10^{-5}$~mol\,m$^{-2}$\,s$^{-1}$, compared with $5.0\times10^{-5}$~mol\,m$^{-2}$\,s$^{-1}$ for reported single-phase Ketton flow and $6.9\times10^{-4}$~mol\,m$^{-2}$\,s$^{-1}$ in batch. Thus, rapid reactant delivery coexisted with restricted mineral access and bypassing under multiphase conditions.

For CO\textsubscript{2} storage in hydrocarbon-bearing carbonate rocks, these results show that bulk $Pe$--$Da$ conditions alone are insufficient to predict dissolution patterns when the pore-scale flow field is strongly heterogeneous. Preferential reactant delivery can amplify existing high-flow pathways and promote channelised dissolution even under a high-$Pe$, low-$Da$ regime. Incorporating flow-field heterogeneity into reactive-transport models is therefore important for predicting channel development and pore-structure evolution during CO\textsubscript{2} injection.

\section*{Acknowledgements}

QM gratefully acknowledges Resource Geophysics Academy, Imperial College London for financial support. The authors also extend their sincere gratitude to Vincenzo Cunsolo for his invaluable assistance in conducting the experiments.

\bibliographystyle{elsarticle-num}
\bibliography{main} 
\end{document}